\documentclass[twocolumn]{emulateapj}
\usepackage[T1]{fontenc}
\usepackage{xcolor}
\usepackage[normalem]{ulem}

\shortauthors{Brittain et al.}
\shorttitle{OH in Herbig Ae/Be Stars}

\begin{document}
\title{A study of ro-vibrational OH Emission from Herbig Ae/Be Stars}

\author{Sean D. Brittain}
\affil{Department of Physics \& Astronomy, 118 Kinard Laboratory, Clemson University, Clemson, SC 29634-0978, USA }
\email{sbritt@clemson.edu}

\author{Joan R. Najita}
\affiliation{National Optical Astronomy Observatory, 950 N. Cherry Ave., Tucson, AZ 85719, USA}

\author{John S. Carr}
\affil{Naval Research Laboratory, Code 7211, Washington, DC 20375, USA}

\author{M{\'a}t{\'e} {\'A}d{\'a}mkovics}
\affil{Department of Astronomy, University of California, Berkeley, CA 94720-3411, USA}

\author{Nickalas Reynolds}
\affil{Department of Physics \& Astronomy, 118 Kinard Laboratory, Clemson University, Clemson, SC 29634-0978, USA }

\begin{abstract}
We present a study of ro-vibrational OH and CO emission from 21 disks around Herbig Ae/Be stars. We find that the OH and CO luminosities are proportional over a wide range of stellar ultraviolet luminosities. The OH and CO line profiles are also similar, indicating that they arise from roughly the same radial region of the disk. The CO and OH emission are both correlated with the far-ultraviolet luminosity of the stars, while the PAH luminosity is correlated with the  longer wavelength ultraviolet luminosity of the stars. Although disk flaring affects the PAH luminosity, it is not a factor in the luminosity of the OH and CO emission. These properties are consistent with models of UV-irradiated disk atmospheres. We also find that the transition disks in our sample, which have large optically thin inner regions, have lower OH and CO luminosities than non-transition disk sources with similar ultraviolet luminosities. This result, while tentative given the small sample size, is consistent with the interpretation that transition disks lack a gaseous disk close to the star. 
\end{abstract}

\section{Introduction}
Herbig AeBe Stars (HAeBes) are young ($\lesssim$10~Myr) intermediate mass (2-8M$_{\odot}$) 
stars whose spectral energy distribution reveals an infrared excess 
\citep{Herbig1960, Waters1998, Vieira2003}. They are 
the precursors to main sequence A-stars many of which have debris disks \citep{Su2005,Su2006} 
and planets (e.g., as evidenced by planets around ``retired'' A-stars (\citealt{Johnson2010});  
the pollution of white dwarfs (\citealt{Zuckerman2010}); and direct imaging of companions  (\citealt{Lagrange2009})).

HAeBes are characterized as either 
Group I  (flared disks) or Group II (self-shadowed 
disks)  sources based on the infrared color of the spectral energy distribution (SED; 
\citealt{Meeus2001};\citealt{Dullemond2010}).
It has been suggested that Group I HAeBes have disks with gaps
\citep[i.e., transitional disks; ][]{Honda2012, Maaskant2013}. If so and if these
gaps are formed by giant planets, this would confirm the 
inference that gas giant planets are common around more massive stars 
(\citealt{Johnson2010}). Thus HAeBes are prime candidates for studying
the signatures of forming supra-Jovian mass companions.

Ro-vibrational emission lines from diatomic molecules such as CO and OH are 
valuable diagnostics of the inner, planet-forming region of disks around young stars.  
The properties of ro-vibrational CO emission from disks surrounding HAeBes have been 
studied extensively \citep[e.g.,][]{Blake2004, Brittain2007, Banzatti2015,vanderplas2015}
However, ro-vibrational OH emission has only been observed in a few HAeBes 
\citep{Mandell2008, Fedele2011, Liskowsky2012}. The observation of OH emission is 
challenging due to the weakness of the features (typically $\lesssim$ 1-3\% of the 
continuum). This is in part because the fundamental ro-vibrational 
transitions of OH are divided over four times as many lines as the fundamental 
ro-vibrational transitions of CO. The total angular momentum of the 
ground electronic state is $\Lambda = 1$ so that it is doubly 
degenerate and each rotational level within the ground electronic state is 
split ({$\Lambda$}-doubling). Thus, for similar 
temperatures and abundances, one would expect the fundamental 
ro-vibrational OH emission lines to be about a factor of four
weaker than the fundamental ro-vibrational CO emission lines. 
However, there has not been a systematic study of 
how the OH and CO emission from HAeBes compare. The goal of this study is 
to characterize the OH emission from a group of HAeBes spanning a wide range 
of spectral types and disk geometries to better understand its relation to the more 
familiar CO emission as well as what determines the abundance and excitation of OH. 
		
Here we present a study of ro-vibrational OH emission from 21
HAeBes. We focus on the $\rm ^2\Pi_{3/2}~P4.5 (1+,1-)$ doublet to 
facilitate comparison with previous studies \citep[e.g.][]{Mandell2008,Fedele2011} 
and because this is the transition with the lowest upper level energy state 
regularly accessible from the ground. At 3,860~cm$^{-1}$ above the ground 
state, the energy of the upper level of this transition is comparable to that of the v=1-0 P19 
CO line (E$^{\prime}$=3,880~cm$^{-1}$).  The spectra analyzed 
in this paper were culled from previous observations
 in the literature \citep{Mandell2008,Fedele2011,Liskowsky2012}, 
 data taken from the archives, and new observations. The 
 sample is roughly evenly divided between Group I and Group II sources spanning
 a broad range of spectral types.  To look at the roles various thermo-chemical process play in the 
excitation of OH emission in disks around HAeBes, we 
explore the relationship between the luminosity of OH emission 
and the disk geometry, ultraviolet luminosity of the star, the
luminosity of polyaromatic hydrocarbon (PAH) emission, and ro-vibrational CO emission. 

\section{Observations}
We present new observations of eight HAeBes in the $L-$band and six HAeBes in the $M-$band (Table 1). 
Seven of the $L-$band observations were obtained at the W. M. Keck Observatory 
with the NIRSPEC echelle spectrograph \citep{McLean1998}. This instrument 
provided a resolving power of $\lambda / \Delta \lambda=25,000$. One 
source, HD~142527, was observed with CRIRES on the European Southern 
Observatory Very Large Telescope (\citealt{Kaeufl2004}). Our team collected 
the data for six of the sources. HD~135344b was taken from the Keck 
Observatory Archive and HD~142527 was taken from the ESO Archive.  Among the $M-$band
observations, three sources (V380~Ori, HD~34282, and CO~Ori) were obtained at Gemini South
Observatory with the PHOENIX echelle spectrograph \citep{Hinkle1998, Hinkle2000, Hinkle2003}. This instrument 
provided a resolving power of $\lambda / \Delta \lambda=50,000$. Three other sources were 
taken from the Keck Observatory Archive (HD~36112, UX~Ori, and BF~Ori). 

Even at the ``blue'' end of the $L-$band where the $\rm v=1-0~P4.5 (1+,1-)$ 
OH doublet falls, there is a significant thermal continuum background upon 
which are superimposed night sky emission lines (mostly water).  The 
intensities of these telluric lines scale with the air mass and column density 
of atmospheric water vapor, which can vary both temporally and spatially 
through the night. To cancel the sky emission, the telescope was nodded 
15$\arcsec$, along the slit axis in an A, B, B, A sequence with each step 
corresponding to 1 minute of integration time.  Combining the scans as 
(A$-$B$-$B+A)/2 canceled the background to first order.  
 
Flats and darks were taken before the grating was moved and used to remove 
systematic effects. The 2-dimensional frames were cleaned of systematically 
hot and dead pixels as well as cosmic ray hits, and were then resampled spatially 
and spectrally so that the spectral and spatial dimensions are orthogonal. The 
CRIRES spectrum was reduced using the ESO CRIRES pipeline
following a similar procedure. 
 
The data were wavelength calibrated by fitting the telluric lines in the spectra with 
the Spectrum Synthesis Program \citep[SSP][]{Kunde1974JQSRT..14..803K}, which 
accesses the  2000HITRAN molecular database \citep{Rothman2003JQSRT..82....5R}. 
Spectra of hot stars were taken at similar airmasses throughout the observations 
for telluric correction. The line fluxes were calculated by scaling the equivalent width to the 
continuum flux inferred from ground based IR photometry. 
The log of the observations is presented in Table 1 and the telluric 
corrected spectra are presented in Figure 1. 

Our full sample of HAeBes have spectral types that range from F6 to B2 (Table 2). 
In addition to the new observations presented here, we include 13 previously 
published OH measurements  \citep{Mandell2008,Fedele2011,Brittain2014} 
and 9 previously published CO measurements \citep{Brittain2007, vanderplas2015} 
of HAeBes (Table 3). 

\section{Results}
The v=1-0 $\rm ^2\Pi_{3/2}~P4.5 (1+,1-)$ OH emission doublet is detected in 4 
of the 8 sources reported in this paper (Fig. 1). This is consistent with the overall 
detection rate of OH reported in the literature 
\citep[6 of 13; Table 3;][]{Mandell2008,Fedele2011,Liskowsky2012}. 

HAeBes are broadly separated into two groups: those with 
flaring disks (Group I) and those with flat disks  \citep[Group II;][]{Meeus2001}. 
What was initially a qualitative description of the infrared portion of the SED, 
was quantified by Acke \& van den Ancker (2004) where systems with 
L$\rm_{NIR}$/L$\rm_{MIR}$ $\leq$ [12]-[60] + 0.9 are Group I and the others
are Group II. A different dividing line was proposed by \citet{vanBoekel2005}
where L$\rm_{NIR}$/L$\rm_{MIR}$ $\leq$ [12]-[60] + 1.5 are Group I sources. 
L$\rm_{NIR}$ is the integrated luminosity from 1--5$\micron$, 
L$\rm_{MIR}$ is the integrated luminosity from 12--60$\micron$, 
and [12]--[60]=--2.5log(L$_{12\micron}$/L$_{60\micron}$).  L$_{12\micron}$ is 
the flux density at 12$\micron$ and L$_{60\micron}$ is the flux density at 60$\micron$. 

The evolving definition of Group I and Group II has resulted in conflicting
designations in the literature. For example, V380~Ori is labeled a Group II star by 
van Boekel et al. (2005) and Group I star by Fedele et al. (2011). Similarly, 
LkH$\alpha$~224 is considered a Group II source by Acke et al. (2004) and 
a Group I source by Acke et al. (2010). Here we separate the sources into Group I and 
Group II using the infrared photometry reported in Table 3 in order to self consistently
define our sample. We adopt the definition of Group I and Group II sources provided 
by \citet{vanBoekel2005}. The near infrared flux was 
taken from 2MASS and WISE. The midinfrared flux was taken from IRAS. 
To integrate the flux from the photometric points, a 100 element array
was created using cubic spline interpolation for both the measured specific flux 
and wavelength. The integral was determined numerically using the trapezoid rule. 

The full sample is comprised of 13 Group I targets and 
8 Group II targets (Fig. 2). This sample includes several disks with 
large inner holes: HD~100546 \citep[inferred from coronographic imagery;][]{Grady2005}, 
HD~135344b\citep[inferred from SED modeling;][]{Brown2007}, HD~142527 
\citep[inferred from SED modeling combined with resolved MIR imagery;][] {Verhoeff2011}, 
HD~179218 \citep[MIR interferometry;][]{menu2015}, and HD~141569 \citep
[inferred from resolved MIR imagery;][]{marsh2002}. HD~142527 has a low 
mass stellar companion orbiting near the inner rim of the circumstellar 
disk \citep[][]{Close2014,Biller2012}.

 The OH doublet is detected in 8 of the 13 
Group I sources (61\%) and 2 of the 8 Group II sources (25\%). OH emission is 
observed in 2 of the 5 transition disk objects (HD~142527 and HD~100546). 
Several of the Group I sources for which OH is detected fall very close to the Group I/II 
boundary, and therefore the difference between the detection rate of OH emission from 
Group I and Group II HAeBes should be interpreted with caution. There is no clear trend 
between the luminosity of the OH emission and infrared colors of these stars (Fig. 2).
 
We plot the luminosity of the v=1-0 $\rm ^2\Pi_{3/2}~P4.5 (1+,1-)$ OH emission doublet 
relative to the v=1-0 P30 CO line (Fig. 3).  The sources for which OH and CO are detected 
are well fit with a power law where 
$\rm log(L_{OH})=(1.01\pm0.02)log(L_{CO}) -(1.05\pm0.02)$.  The typical ratio of the luminosity 
of the CO to the OH is 11.0$\pm$0.2. None of the non-detections of OH emission 
are sensitive enough to place an upper limit on the OH/CO ratio lower than the line fit to the 
detections. If the temperature and emitting area of the OH and CO gas are similar and 
the gas is optically thin, then the abundance of the OH gas is comparable to CO in 
systems spanning several dex in UV luminosity (see Table 3). The transition disks  
are clustered among the weakest OH and CO emitters in the sample (Figs. 3, 5, and 6).

For five of the sources in our sample, we have resolved OH and CO emission lines. The 
line profiles of the OH and CO emission lines are consistent to within 
the signal to noise for four of the stars (though the S/N of HD~142527 is too low to say 
anything definitive; Fig. 4). The exception is HD~100546 where the OH and CO line 
profiles are not consistent \citep[see also][]{Liskowsky2012}. While the widths of the HD~100546 lines
are similar, the shapes of the asymmetries are different. Previous studies of the CO emission
show that the the v=1-0 lines are comprised of two components -- a symmetric line arising from
the circumstellar disk extending from $\sim$14~AU to beyond 50~AU and a variable component 
arising from a localized source of emission in a Keplerian orbit near the inner edge of the 
circumstellar disk \citep{Brittain2014}. 

To explore the role the stellar UV field plays in producing the observed 
molecular emission, we compare the luminosity of the OH, CO, and PAH emission 
to the ultraviolet luminosity of the HAeBes 
(Figs. 5-7). Most of the sources were observed 
with IUE \citep[Table 3; ][]{Valenti2000, Valenti2003}. 
For sources not observed with IUE, we determined the spectral type of the star and 
calculated the UV luminosity from a matching stellar template (see Table 3 for the template stars). 
One star, HD~142527 (SpT F6), has no template or UV observation and is left out of this analysis. 
Stars with a spectral type later than A5 may have a 
UV flux dominated by the accretion luminosity of the star, thus
the UV luminosity reported for these objects should be taken as a lower limit. We consider 
the short wavelength portion of the UV spectrum (1300\AA~$\leq \lambda \leq$~1850\AA ; 
L$\rm_{UVS}$) separately from the long wavelength portion (2430\AA~$ \leq \lambda 
\leq$~3200\AA ; L$\rm_{UVL}$). The radiation at the short wavelength 
end of the spectrum enhances OH emission both by photodissociating 
H$_2$O to produce OH, possibly in an excited state,
 and by heating the gas. The longer wavelength radiation excites PAHs. 

In figure 5, we plot the luminosity of the OH doublet as a function 
of the short wavelength ultraviolet luminosity. The sources that are 
brightest at ultraviolet wavelengths also reveal the brightest OH 
emission lines. The stars with transition disks have weaker OH 
emission than HAeBes of the same L$\rm_{UVS}$.  To quantify 
the correlation between the logarithm of the OH 
luminosity and the logarithm of the short wavelength luminosity we apply 
Kendall's $\tau$ correlation test using ASURV Rev 1.2 (Isobe \& Feigelson 
1990; LaValley, Isobe \& Feigelson 1992), which implements the methods 
presented in (Isobe,Feigelson \& Nelson 1986). This test finds that there 
is a 1.5\% chance that a correlation is not present in the data. We use the 
Buckley-James regression test using ASURV Rev 1.2 to find the best 
least-squares fit to the data and find $\rm log(L_{OH}/10^{-5}L_{\sun})
=(0.45\pm0.25)log(L_{UVS}/L_{\sun})-0.93 $.

One outlier in this data set is the transition disk HD~141569 (source 15). 
HD~141569  is a low mass transition disk with a large inner hole and 
cool molecular gas (e.g., \citealt*{Zuckerman1995}; \citealt{Brittain2003, Dent2005}). 
This source has a particularly low gas \citep{Thi2014} and dust mass 
\citep{Dent2005} and bridges the gap between gas rich transition disks and debris disks 
mostly emptied of their primordial gas.  If the OH and CO in the 
system are roughly the same temperature ($\sim$200~K; Brittain et al. 2007), it is unlikely that the upper 
states of the v=1-0 $\rm ^2\Pi_{3/2}~P4.5 (1+,1-)$ OH doublet will be populated at the 
same level as the other systems with much warmer gas. Unfortunately, the lower-J OH 
lines are generally not accessible from the ground except under 
exceptionally dry conditions. If we set aside HD~141569 from our 
statistical analysis, the probability that the data are not correlated 
does not change substantially (1\%), and the fit to the data from 
the Buckley-James method changes to $\rm log(L_{OH}/10^{-5}
L_{\sun})=(0.48\pm0.24)log(L_{UVS}/L_{\sun})-0.51$.

In figure 6, we plot the the luminosity of the CO emission as a function 
of the short wavelength ultraviolet luminosity and see a similar trend. 
The stars with transition disks also show systematically lower CO 
emission than their classical counterparts with comparable L$\rm_{UVS}$.
The correlation test by the 
generalized Kendall's $\tau$ indicates that the probability that the data 
are not correlated is 0.07\% if we set aside HD~141569. The linear 
regression using the Buckley-James method indicates that $\rm 
log(L_{CO}/10^{-5}L_{\sun})=(0.68\pm0.17)log(L_{UVS}/L_{\sun})+0.47 $. 

Finally, we consider the relationship between the log of the luminosity 
of the PAHs and the log of the UV luminosity (Fig. 7). 
While the trend is similar, 
there is more scatter and the transition disk objects are not systematically 
fainter than the other sources. However, the luminosities of the PAH emission from 
the Group II sources is lower than those from the Group I sources with the same long-wavelength UV 
luminosity. The probability that the quantities are not 
correlated is 5\% and the linear regression by the Buckley-James method 
indicates that $\rm log(L_{PAH}/10^{-2}L_{\sun})=(0.37\pm0.12)log(L_{UVL}/L_{\sun})+0.61 $.

\section{Discussion and Conclusions}
The disks surrounding HAeBes are gas rich as evidenced by their high 
accretion rates \citep{Fairlamb2015, Mendigutia2015}. However, HAeBes are powerful 
UV emitters, making protoplanetary disks challenging environments for 
the synthesis and survival of prebiotic molecules, particularly when 
grains, the usual UV absorbers, have settled out of disk atmospheres. 
Recent studies of T Tauri disks, the lower mass counterparts of 
HAeBes, show how water synthesized in disk 
atmospheres absorbs stellar UV through photodissociation, yielding 
OH \citep{Bethell2009, Du2014, Adamkovics2014}.

The trends observed in Figures 3, 5, 6, and 7 can be explained 
simply in this picture of a disk irradiated by stellar UV photons. 
In dissociating water at the disk surface, the UV not only enhances 
the abundance of OH there, but it also heats the disk surface \citep[][]{Du2014, 
Adamkovics2014, Adamkovics2015, Glassgold2015}.  

In models of UV-irradiated disk atmospheres \citep[e.g.,][]{Adamkovics2015}, 
CO and OH are both abundant in the warm surface region of the 
disk atmosphere (at $\sim$1000K), which is heated by UVS (FUV 
continuum and Ly$\alpha$) through photodissociation and photochemical 
pathways. In accordance with this picture, we find that the CO and OH 
line profiles are consistent with each other (Fig. 4), hinting that the gas 
arises from the same region of the disk, as expected from models. 

The excitation mechanism of the OH emission discussed here is uncertain. 
In principle, OH can be UV fluoresced similarly to CO as OH possesses 
bound excited electronic states. However, collisions can quench UV fluorescence
such that the final population of the vibrational levels of OH are determined collisionally. 
When collisional cross sections for vibrational excitation of OH with partner species such as H,
H$_2$, and He are available, detailed studies OH excitation can be carried out. 
In the meantime, our data hints that the OH and CO excitation mechanism are 
related, as the ratio of their fluxes are constant over a wide range of disk 
geometries and UV luminosities (Fig. 3).

Group I disks are flared while Group II disks are flat \citep{Meeus2001}, 
thus Group I disks intercept a larger fraction of the stellar flux. Because of this, 
one might expect that the ro-vibrational emission of molecules would be brighter 
among Group I disks than among Group II disks. This difference is not observed (Fig. 2). 
Nor do we find a difference in the relationship between OH and the UV 
luminosity  or CO and the UV luminosity between Group I HAeBes and 
Group II HAeBes. One possibility is that this grouping conflates 
varying degrees of disk clearing and flaring
\citep[e.g.,][]{menu2015} obscuring any distinction in
luminosity of ro-vibrational emission lines between Group I and Group II sources.

We have identified five systems 
in our sample with cleared inner disks. 
Compared to the ``normal'' HAeBes, transition disks are missing 
the warm disk close to the star, so they are expected to have 
systematically fainter OH and CO emission relative to full disks
(Figs. 3, 5, \& 6). In contrast to the CO and OH emission, which 
depends on the intensity of incident UVS photons, PAH emission 
can be generated by UVL photons that are absorbed at any distance 
from the star.  As long as the disk 
intercepts the long wavelength ultraviolet radiation at some radius, 
PAH emission can be produced.  As a result, PAH emission should depend less sensitively 
on whether the inner disk is missing or not. Accordingly, the transition disk 
objects are not systematically offset relative to the full disks (Fig. 7). 
However, the amount of disk flaring will affect the PAH luminosity. 
As a result, the less flared Group II sources have systematically 
fainter PAH flux for the same long-wavelength UV luminosity. 

Further work can better quantify the relation of ro-vibrational OH emission to ro-vibrational CO emission.
We did not measure the rotational temperature of the OH for the stars 
in our sample, nor do we have the data necessary to explore the excitation mechanism of the OH.
 For most of the sources where OH is detected, only the P4.5 doublet was targeted. 
In a few sources where there was a detection of OH, there are no corresponding CO 
observations. If the OH and CO are co-radially located but at different temperatures, this will 
suggest that we are probing gas at different depths into the disk. The data do not show 
evidence of enhanced OH emission concentrated at the inner rim of disks
with holes, however, higher resolution and signal to noise data of the transition disk objects
are necessary to draw firm conclusions. 
The availability of new large format detectors on high 
resolution NIR  spectrographs such as iSHELL on NASA's Infrared Telescope Facility and the 
refurbished CRIRES on ESO's VLT will enable such studies. 

\acknowledgments
Some of the data presented herein were obtained at the W.M. 
Keck Observatory, which is operated as a scientific partnership 
among the California Institute of Technology, the University of 
California and the National Aeronautics and Space Administration. 
The Observatory was made possible by the generous financial 
support of the W.M. Keck Foundation. The authors wish to 
recognize and acknowledge the very significant cultural role and 
reverence that the summit of Mauna Kea has always had within 
the indigenous Hawaiian community.  We are most fortunate to 
have the opportunity to conduct observations from this mountain.  
This research has also made use of the Keck Observatory Archive, 
which is operated by the W. M. Keck Observatory and 
the NASA Exoplanet Science Institute, under contract 
with the National Aeronautics and Space Administration. Based also
in part on data obtained from the ESO Science Archive Facility.  Based 
also in part on observations obtained at the Gemini Observatory 
(GS-2005B-C-2) which 
is operated by the Association of Universities for Research in Astronomy, 
Inc., under a cooperative agreement with the NSF on behalf of the Gemini 
partnership: the National Science Foundation (United States), the National 
Research Council (Canada), CONICYT (Chile), Ministerio de Ciencia, 
Tecnolog\'{i}a e Innovaci\'{o}n Productiva (Argentina), and Minist\'{e}rio 
da Ci\^{e}ncia, Tecnologia e Inova\c{c}\~{a}o (Brazil). This paper is based
in part on observations obtained with the Phoenix infrared
spectrograph, developed and operated by the National Optical Astronomy
Observatory.
S.D.B.  acknowledges 
support for this work from the National Science Foundation 
under grant number  AST-0954811. Basic research in 
infrared astronomy at the Naval Research Laboratory is 
supported by 6.1 base funding. 

{\it Facilities:} \facility{Keck (NIRSPEC)}, \facility{Very Large Telescope:UT1 (CRIRES)}, \facility{Gemini South (PHOENIX)}

\nocite{*}
\bibliographystyle{apj}
\bibliography{ohpapercite_clean}

\clearpage

\begin{deluxetable}{lllll}
\tablenum{1}
\tablecaption{Log of Observations}

\tablehead{ 	\colhead{Star}	&	\colhead{Date Obs}	&	\colhead{Airmass}	&	\colhead{Int Time}	\\
			\colhead{}		&	\colhead{}			&	\colhead{}			&	\colhead{m}	}
\startdata
\sidehead{$L-$band data}								
HD~179218		&	September~13,~2014	&	1.02	&	48	\\
V380~Ori			&	September~13,~2014	&	1.46	&	44	\\
HD~135344b		&	July~13,~2009			&	1.84	&	16	\\
LkH$\alpha$~224	&	September~13,~2014	&	1.09	&	56	\\
HD~142527		&	May~29,~2013			&	1.05	&	10	\\
Elias~3-1			&	September~13,~2014	&	1.24	&	76	\\
HD~141569		&	April~8,~2014			&	1.16	&	32	\\
HD~190073		&	September~13,~2014	&	1.31	&	40	\\
\sidehead{$M-$band data}	
V380~Ori			&	January 13, 2006		&	1.65	& 	16	\\
CO~Ori			&	January 13, 2006		&	1.91	&	20	\\
HD~34282		&	January 14, 2006		&	1.07	&	40	
\enddata
\end{deluxetable}

\clearpage

\begin{deluxetable}{lllllll}
\tablenum{2}
\tablecaption{Stellar Parameters}

\tablehead{
\colhead{Number}	&	\colhead{Star}	&	\colhead{Alt Name}	&	\colhead{SpT\tablenotemark{a}}	&	\colhead{d\tablenotemark{i}}			&	\colhead{Group}	&	\colhead{log(T$\rm_{eff}$)\tablenotemark{o}}	\\
\colhead{}			&	\colhead{}		&	\colhead{}			&	\colhead{}			&	\colhead{pc}			&	\colhead{}			&	\colhead{K}			}
\startdata															
1	&	HD~76534				&	OU~Vel		&	B2					&	870\tablenotemark{j}				& I		& 4.28	\\
2	&	HD~259431				&	MWC~147	&	B5					&	170$\pm$40					& I		& 4.15	\\
3	&	HD~250550				&	V1307~Ori	&	B7					&	1176$\pm$1993				& I		& 4.11	\\
4	&	HD~100546				&	KR~Mus		&	B9					&	97$\pm$4						& I		& 4.04	\\
5	&	HD~31293				&	AB~Aur		&	A0					&	140$\pm$20					& I		& 3.99	\\
6	&	HD~179218				&	MWC~614	&	A0					&	250$\pm$40					& I		& 3.99	\\
7	&	MWC~765				&	V380~Ori		&	A1					&	450\tablenotemark{k}			& I		& 3.97	\\
8	&	HD~36112				&	MWC~758	&	A8\tablenotemark{b}		&	280$\pm$70					& I		& 3.87	\\
9	&	HD~135344b				&	SAO~206462	&	F4\tablenotemark{c}		&	140\tablenotemark{c}			& I 		& 3.83	\\
10	&	LkH$\alpha$~224			&	V1686~Cyg	&	F9/A4\tablenotemark{d}	&	980[13]						& I		& 3.92	\\
11	&	HD~142527				&	\nodata		&	F6\tablenotemark{e}		&	140$\pm$20\tablenotemark{e}		& I		& 3.82\tablenotemark{e}	\\
12	&	HD~45677				&	FS~Cma		&	B2					&	280$\pm$70					& II		& 4.33	\\
13	&	HD~85567				&	V596~Car		&	B2\tablenotemark{f}		&	770\tablenotemark{f}				& II		& 4.33	\\
14	&	Elias~3-1					&	V892~Tau		&	B8.5-A0\tablenotemark{g}	&	140\tablenotemark{g}			& I		& 4.02	\\
15	&	HD~141569				&	\nodata		&	B9.5					&	103$\pm$8					& II		& 4.01	\\
16	&	HD~98922				&	\nodata		&	B9/A2				&	1200$\pm$500					& II		& 3.95	\\
17	&	HD~34282				&	V1366~Ori	&	A3\tablenotemark{h}		&	350\tablenotemark{h}			& I		& 3.93	\\
18	&	HD~190073				&	V1295~Aql	&	A2					&	767\tablenotemark{m}			& II		& 3.95	\\
19	&	HD~293782				&	UX~Ori		&	A3					&	400\tablenotemark{k}			& II		& 3.93	\\
20	&	\nodata					&	BF~Ori		&	A5					&	450\tablenotemark{k}			& II		& 3.91	\\
21	&	\nodata					&	CO~Ori		&	F7					&	450\tablenotemark{l}				& II		& 3.80	\\
22	&	\tablenotemark{p}HD~37776	&	\nodata		&	B2					&	330$\pm$60					& \nodata	& \nodata	\\
23	&	\tablenotemark{p}HD~196519	&	\nodata		&	B8					&	241$\pm$16					& \nodata	& \nodata	\\
24	&	\tablenotemark{p}HD~172167	&	\nodata		&	A0					&	7.7$\pm$0.02					& \nodata	& \nodata	\\
25	&	\tablenotemark{p}HD~102647	&	\nodata		&	A3					&	11$\pm$0.06					& \nodata	& \nodata	\\
26	&	\tablenotemark{p}HD~13041	&	\nodata		&	A4					&	58.7$\pm$0.9					& \nodata	& \nodata	\\
27	&	\tablenotemark{p}HD~11636	&	\nodata		&	A5					&	18$\pm$0.2					& \nodata	& \nodata	\\
28	&	\tablenotemark{p}HD~28910	&	\nodata		&	A8					&	49$\pm$1						& \nodata	& \nodata	\\
29	&	\tablenotemark{p}HD~8799	&	\nodata		&	F4					&	25$\pm$0.3					& \nodata	& \nodata	
\enddata		
\tablenotetext{a}{The spectral types are adopted from Valenti et al. 2003 unless o
therwise noted. These authors adopted the spectral type from either the 
Herbig-Bell Catalog or Th\'{e} et al. (1994) unless they found a large discrepancy 
between the UV spectrum of the star and the template spectrum. In such cases they adopted 
spectral type of the best fitting template.} 
\tablenotetext{b}{Vieira et al. 2003}
\tablenotetext{c}{Dunkin et al. 1997; Grady et al. 2009}
\tablenotetext{d}{A4, Mora et al. 2001; F9, Hernandez et al. 2004}
\tablenotetext{e}{Mendigutia et al. 2014}
\tablenotetext{f}{Hales et al. 2014}
\tablenotetext{g}{Mooley et al. 2013}
\tablenotetext{h}{Merin et al. 2004}
\tablenotetext{i}{from van Leewen et al. 2007 unless otherwise noted}
\tablenotetext{j}{Herbst 1975}
\tablenotetext{k}{distance inferred by Blondel \& Tjin A Djie 2006 based on cluster membership}
\tablenotetext{l}{distance based on membership in l Ori, and cluster distance determined by Dolan \& Mathieu 2001}
\tablenotetext{m}{Montesino et al. 2009}
\tablenotetext{n}{distance inferred from membership in star forming region 2 Cyg and distance of 980pc - Shevchenko, Ibragimov, \& Chernysheva 1991}
\tablenotetext{o}{Effective temperatures converted from the intrinsic B-V color using the relationship described in Flower (1996) unless otherwise noted.}
\tablenotetext{p}{Template stars used to estimate the UV luminosity of stars without IUE data}

\end{deluxetable}

\clearpage

\begin{deluxetable}{lllllllllll}
\tablenum{3}
\tablecolumns{11}
\tablewidth{0pt}
\tablecaption{Circumstellar Data}

\tablehead{
\colhead{No.} &	 \colhead{Star}  &\colhead{[12]-[60]} & \colhead{L$\rm_{NIR}$/L$\rm_{MIR}$} & \colhead{L$\rm_{UV}$\tablenotemark{a}} & \colhead{L$\rm_{UVS}$\tablenotemark{b} } & \colhead{L$\rm_{UVL}$\tablenotemark{c} } &	\colhead{L$\rm_{PAH}$} & \colhead{L$\rm_{PAH}$} & \colhead{L$\rm_{CO}$ }   & \colhead{L$\rm_{OH}$} 		\\
\colhead{} & \colhead{} & \colhead{} & \colhead{} & \colhead{}	& \colhead{}	& \colhead{}	& \colhead{7.8\micron }	& \colhead{11.3\micron }	& \colhead{(1,0)P30}		& \colhead{(1,0)P4.5(+1,-1)} \\
\colhead{} & \colhead{} & \colhead{} & \colhead{} & \colhead{L$_{\sun}$}	& \colhead{L$_{\sun}$}	& \colhead{L$_{\sun}$}	& \colhead{10$^{-2}$L$_{\sun}$}	& \colhead{10$^{-2}$L$_{\sun}$}	& \colhead{10$^{-5}$L$_{\sun}$}	& 10$^{-5}$L$_{\sun}$ } 

\startdata
1 &	HD~76534		&	6.08	&	2.13	&	1430\tablenotemark{d}	&	1043\tablenotemark{d}	&	166\tablenotemark{d}	&	\nodata		&	\nodata		&	\nodata						&	$\leq$3.5\tablenotemark{g}		\\	
2 &	HD~259431		&	2.35	&	1.10	&	15.1					&	10.4					&	3.5					&	\nodata		&	2.86			&	16$\pm$3\tablenotemark{e}		&	1.1$\pm$0.1\tablenotemark{g}		\\	
3 &	HD~250550		&	0.51	&	1.75	&	157					&	93					&	48					&	$\leq$25.7		&	$\leq$3.86		&	77$\pm$3\tablenotemark{e}		&	6.7$\pm$0.5\tablenotemark{g}		\\	
4 &	HD~100546		&	1.00	&	0.34	&	8.5					&	5.9					&	2					&	9.31			&	11.3			&	4.2$\pm$0.2\tablenotemark{f}		&	0.26$\pm$0.03\tablenotemark{h}	\\	
5 &	HD~31293		&	1.47	&	2.49	&	7.8					&	4.2					&	3.3					&	10.8			&	0.66			&	12.0$\pm$0.8\tablenotemark{e}		&	0.4$\pm$0.1\tablenotemark{i}		\\	
6 &	HD~179218		&	0.27	&	1.01	&	21.7					&	13.5					&	6.8					&	28.4			&	2.78			&	3.0$\pm$0.2\tablenotemark{f}		&	$\leq$0.6\tablenotemark{j}			\\	
7 &	MWC~765		&	2.36	&	1.30	&	8.6					&	5.7					&	3.1					&	\nodata		&	\nodata		&	21$\pm$2						&	4.9$\pm$0.7\tablenotemark{g}		\\	
8 &	HD~36112		&	1.75	&	2.70	&	0.48\tablenotemark{d}	&	0.005\tablenotemark{d}	&	0.94\tablenotemark{d}	&	$\leq$2.42		&	$\leq$0.48		&	12.3$\pm$1.4\tablenotemark{f}		&	0.9$\pm$0.2\tablenotemark{i}		\\	
9 &	HD~135344b		&	3.02	&	3.45	&	0.05\tablenotemark{d}	&	0.001\tablenotemark{d}	&	0.19\tablenotemark{d}	&	0.85			&	0.16			&	1.3$\pm$0.2					&	$\leq$0.3\tablenotemark{j}			\\	
10 &	LkH$\alpha$~224	&	2.23	&	0.46	&	1.8\tablenotemark{d}		&	0.6\tablenotemark{d}		&	1.20\tablenotemark{d}	&	$\leq$71.7		&	$\leq$2.08		&	\nodata						&	11.8$\pm$2.7\tablenotemark{j}		\\	
11 &	HD~142527		&	2.51	&	1.87	&	\nodata				&	\nodata				&	\nodata				&	\nodata		&	\nodata		&	2.8$\pm$0.3					&	0.1$\pm$0.1\tablenotemark{k}		\\	
12 &	HD~45677		&	-1.93	&	0.85	&	1430\tablenotemark{d}	&	1043\tablenotemark{d}	&	166\tablenotemark{d}	&	\nodata		&	\nodata		&	\nodata						&	$\leq$2.4\tablenotemark{g}		\\	
13 &	HD~85567		&	-1.65	&	6.78	&	1430\tablenotemark{d}	&	1043\tablenotemark{d}	&	166\tablenotemark{d}	&	42.6			&	7.17			&	\nodata						&	17.9$\pm$5.4\tablenotemark{g}		\\	
14 &	Elias~3-1			&	0.85	&	0.29	&	8.3\tablenotemark{d}		&	5.1					&	2.6					&	7.47			&	1.52			&	\nodata						&	$\leq$0.2\tablenotemark{j}			\\	
15 &	HD~141569		&	2.51	&	10.7	&	8.3\tablenotemark{d}		&	5.1					&	2.6					&	0.38			&	0.03			&	0.09$\pm$0.01\tablenotemark{e}	&	$\leq$0.02	\tablenotemark{j}		\\
16 &	HD~98922		&	-2.03	&	4.96	&	764					&	394					&	318					&	\nodata		&	73.8			&	172$\pm$9\tablenotemark{f}		&	$\leq$47\tablenotemark{g}			\\	
17 &	HD~34282		&	2.93	&	1.74	&	1.2					&	0.45					&	0.75					&	12.8			&	2.1			&	$\leq$3						&	$\leq$0.8\tablenotemark{g}		\\	
18 &	HD~190073		&	-1.43	&	6.72	&	35.6					&	15.8					&	25.8					&	$\leq$5.47		&	$\leq$3.65		&	62.3$\pm$2\tablenotemark{f}		&	6.5$\pm$0.5\tablenotemark{j}		\\	
19 &	UX~Ori			&	0.07	&	2.63	&	0.51					&	0.23					&	0.32					&	$\leq$1.15		&	$\leq$0.36		&	2.2$\pm$0.2					&	$\leq$1.5\tablenotemark{g}		\\	
20 & BF~Ori			&	1.02	&	4.5	&	0.6					&	0.14					&	0.57					&	$\leq$1.48		&	$\leq$0.56		&	$\leq$1.4						&	$\leq$0.6\tablenotemark{g}		\\	
21 & CO~Ori			&	0.27	&	9.81	&	2.3					&	0.6					&	4.1					&	\nodata		&	\nodata		&	1.7$\pm$0.9					&	$\leq$0.6\tablenotemark{g}		
\enddata

\tablenotetext{a}{Integrated UV flux from 1300\AA\ - 2430\AA . From IUE spectra presented by Valenti et al. (2000, 2003) unless otherwise noted}
\tablenotetext{b}{Integrated UV flux from 1300\AA\ - 1850\AA . From IUE spectra presented by Valenti et al. (2000, 2003) unless otherwise noted}
\tablenotetext{c}{Integrated UV flux from 2430\AA\ - 3200\AA . From IUE spectra presented by Valenti et al. (2000, 2003) unless otherwise noted}
\tablenotetext{d}{UV flux inferred from a template star with a matching spectral type from Valenti et al. (2003).}
\tablenotetext{e}{Brittain et al. 2007}
\tablenotetext{f}{van der Plas et al. 2015}
\tablenotetext{g}{Fedele et al. 2011}
\tablenotetext{h}{Brittain et al. 2014}
\tablenotetext{i}{Mandel et al. 2008}
\tablenotetext{j}{This work}

\end{deluxetable}

\clearpage

\begin{figure}
\centering
\includegraphics[width=6in]{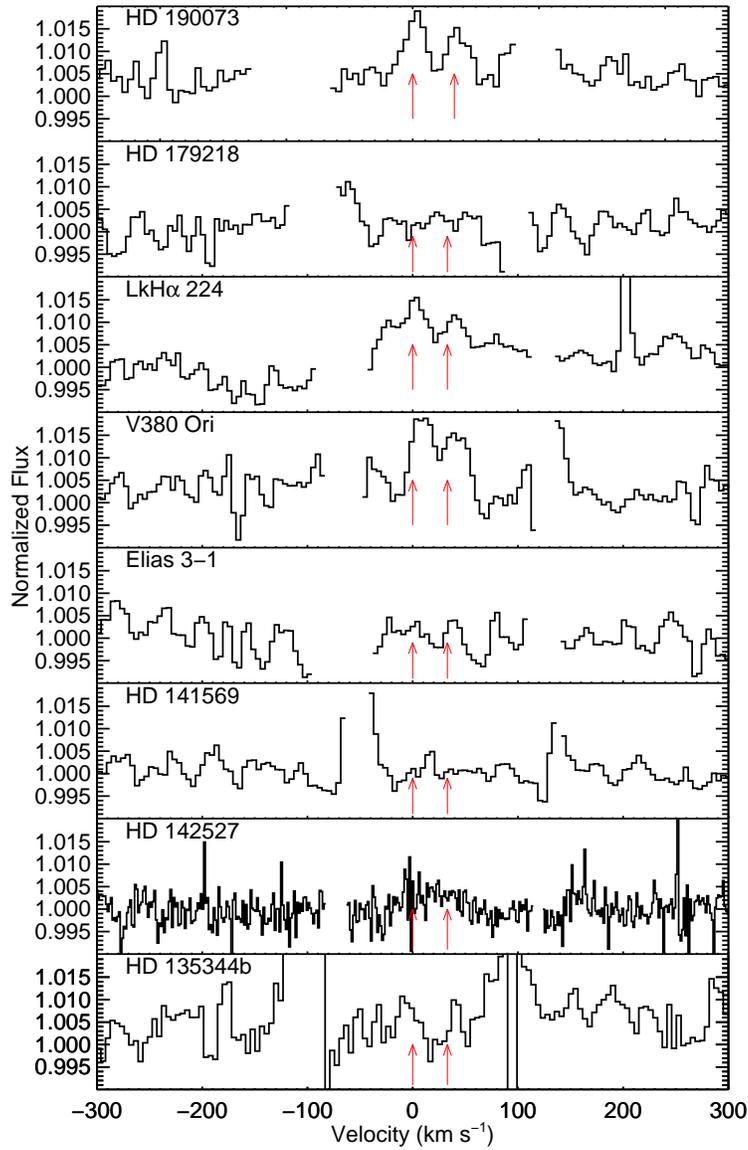}
\caption{Spectra covering the location of the OH v=1--0 P4.5 doublet. OH emission 
is detected in HD~190073, Lk H$\alpha$~224, V380~Ori, and HD~142527. The spectra
are plotted in the rest-frame of the star and the expected location of the OH doublet is marked
with a pair of red arrows.}
\label{ }
\end{figure}

\begin{figure}
\begin{center}
\includegraphics[width=4in, angle=90]{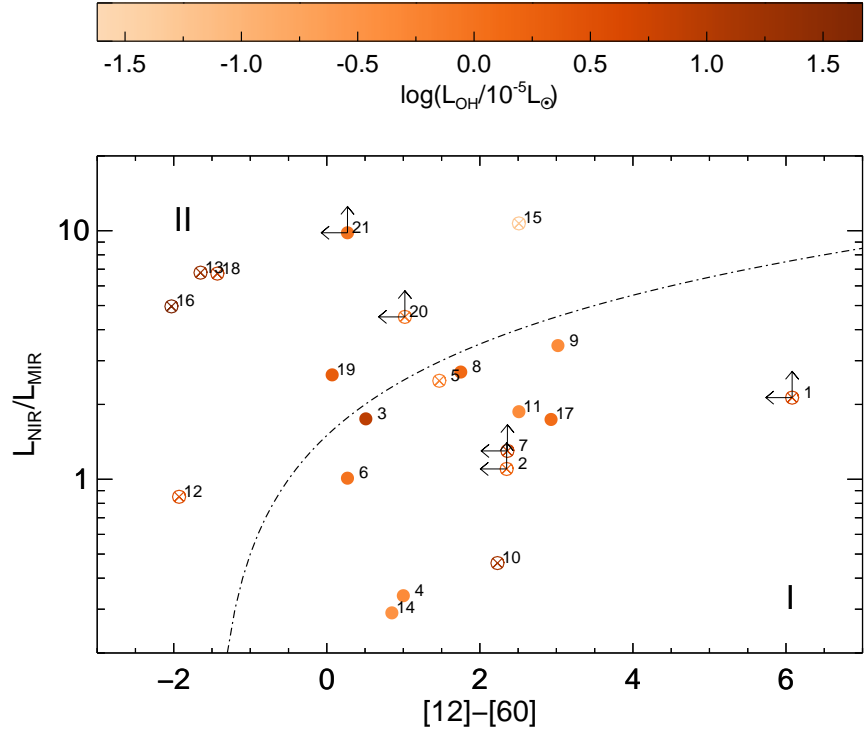} 
\caption{Group selection of sources. Herbig Ae/Be stars are separated into 
		two groups on the basis of their IR colors. Group I sources have 
		L$\rm_{NIR}$/L$\rm_{IR}$ $\leq$ [12]--[60]+1.5 (e.g., van Boekel 
		et al. 2005). The dot-dashed line separates the Group I and Group II sources. 
		The filled sources are sources for which OH was detected and the open circles 
		are non-detections. The logarithm of the luminosity of the observed line or its 
		upper limit is indicated by the color scale. Several of our targets fall near the 
		Group I/Group II boundary.  There is no apparent trend between the luminosity 
		of the OH emission and the group classification. }
\label{ }
\end{center}
\end{figure}

\begin{figure}
\begin{center}
\includegraphics[width=4.5in, angle=-90]{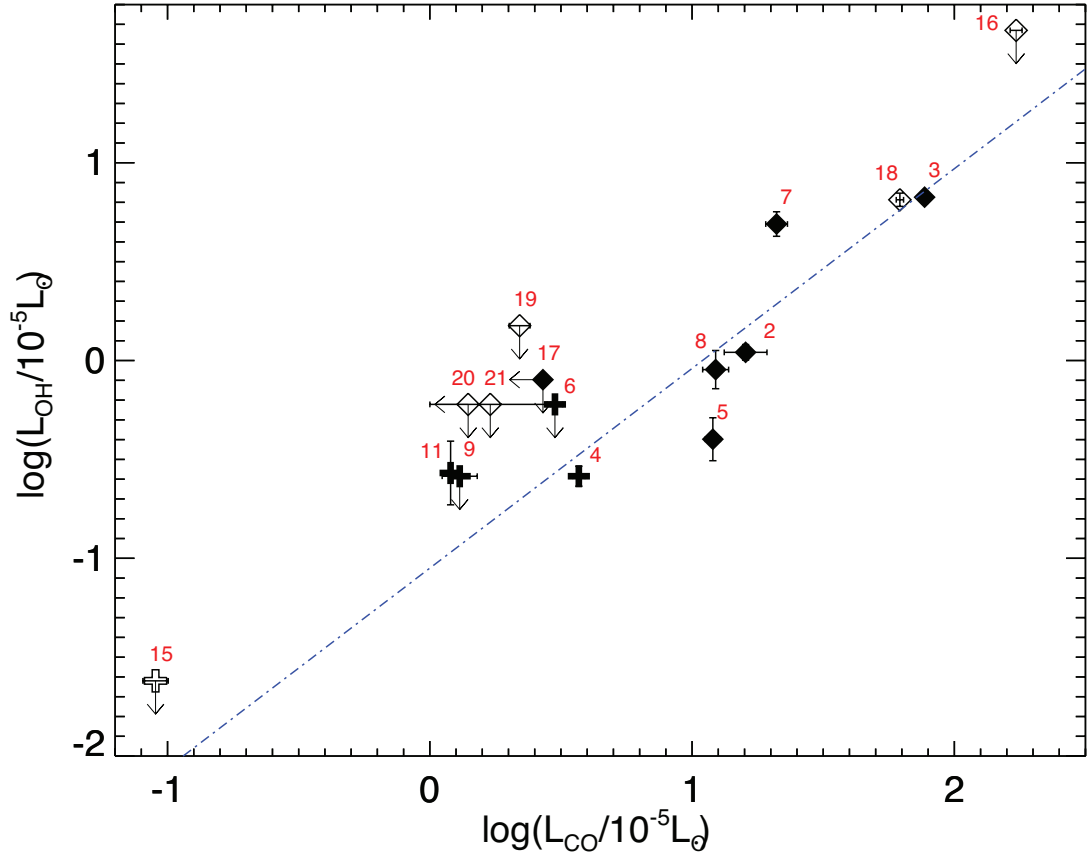} 
\caption{The logarithm of the luminosity of the v=1--0 $\rm ^2\Pi_{3/2}~P4.5 (1+,1-)$ 
OH doublet is plotted versus the logarithm of the v=1--0 P30 CO line. Group I 
Herbig Ae/Be stars are marked with filled points. Group II Herbig Ae/Be stars are
marked with open symbols.  The disks for which holes have 
been identified are indicated with ``plus'' signs, while those without holes are indicated
with diamonds. A linear least square fit to the detections 
is plotted with a blue dot-dashed line. None of the upper limits on the non-detection 
of OH emission falls below this trend. OH emission is only detected in two transition disk
sources, HD~142527 and HD~100546. The luminosity 
 of OH emission tends to be weaker for transitional disks compared to disks without inner holes.}
\label{ }
\end{center}
\end{figure}

\begin{figure}
\includegraphics[width=3in, angle=0]{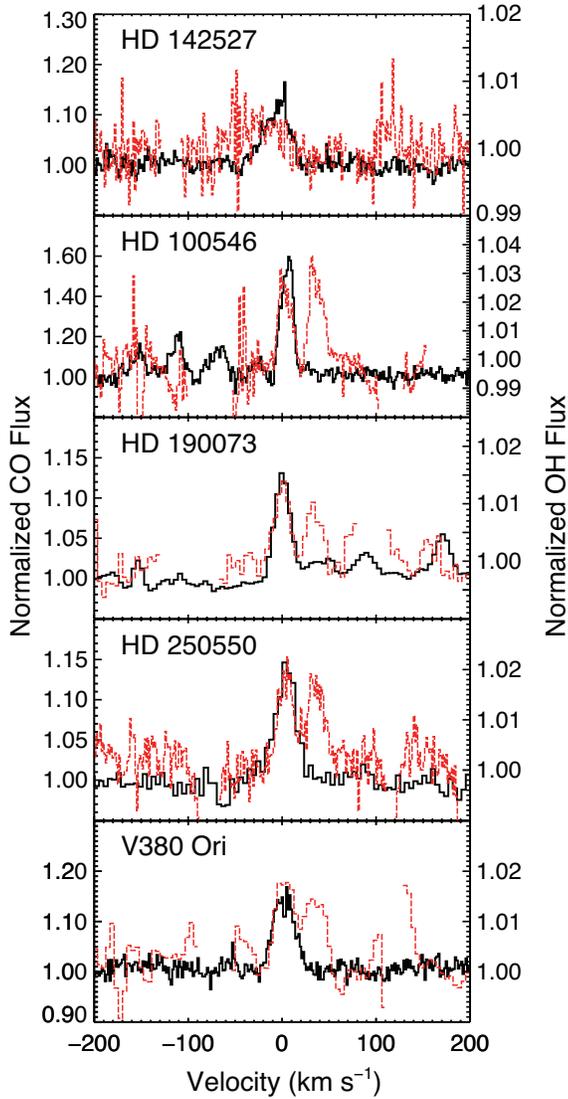} 
\caption{Comparison of a high-J v=1--0 CO line with the v=1--0 P4.5(+1,--1) 
OH doublet. For HD~100546, and V380~Ori, the v=1--0 P26 CO line is 
compared with the v=1-0 P4.5(+1) OH line. For HD~190073 
and HD~250550 the v=1--0 P30 CO line is compared with the =1-0 P4.5(+1) OH line.
For HD~142527 the v=1--0 P26 CO line is compared to the v=1-0 P4.5(--1) line.
In each panel, the CO spectrum is plotted with a black line, 
the OH spectrum is plotted with a red dashed line. There is no significant difference between the line profile 
of the OH and CO lines for HD~190073, HD~250550, or V380~Ori. In the 
case of HD~100546, the shape of the lines are different (see Liskowsky et al. 2012). 
The S/N of the OH line profile for HD142527 is too low to say anything definitive.}
\label{ }
\end{figure}

\begin{figure}[t]
\begin{center}
\includegraphics[width=4.5in, angle=-90]{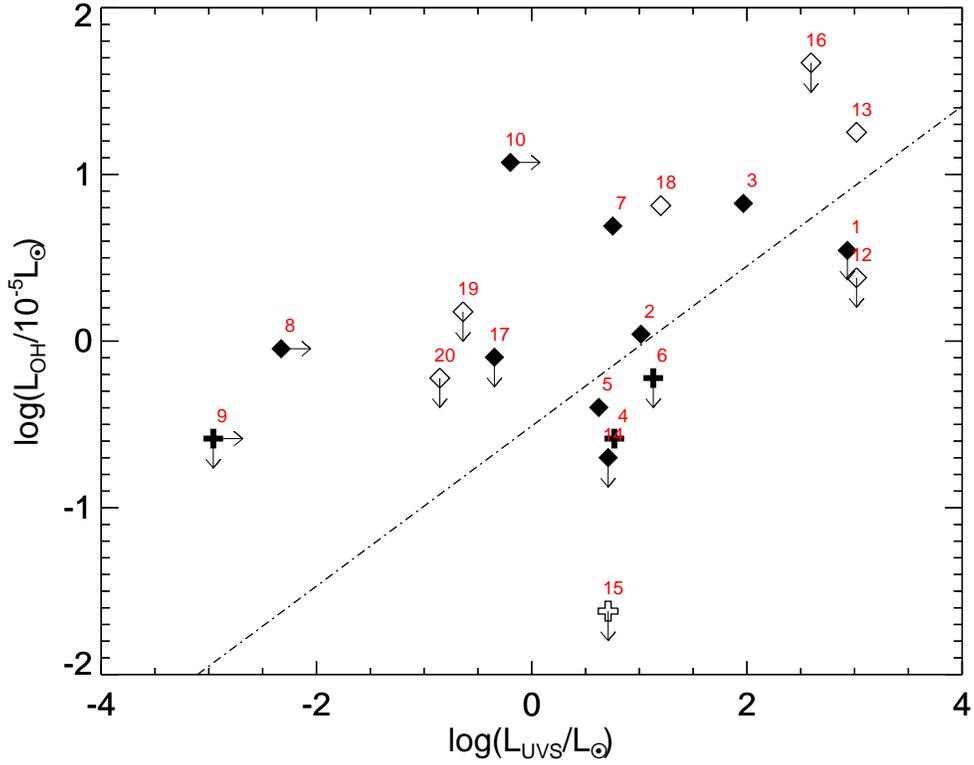} 
\caption{The logarithm of the luminosity of the v=1--0 $\rm ^2\Pi_{3/2}~P4.5 (1+,1-)$ 
OH doublet is plotted versus the logarithm of the ultraviolet luminosity integrated from 
1300\AA\ to 1850\AA\  (L$\rm_{UVS}$). The same symbols are repeated from figure 3.
Several of the stars do not have direct UV flux measurements. 
The UV luminosity of  these stars was inferred from main-sequence template stars. 
The accretion luminosity of stars with a spectral-type later 
 than A5 may provide a significant fraction of the flux that is unaccounted for by this 
 measurement, thus they are noted as lower limits. The transition disks in the sample are systematically
 offset from the broader sample. The result of the Buckley-James regression is plotted with the
 dot-dashed line $\rm 0.48log(L_{UVS}/L_{\sun}) -0.51$ }
\label{ }
\end{center}
\end{figure}

\begin{figure}[t]
\begin{center}
\includegraphics[width=4.5in, angle=-90]{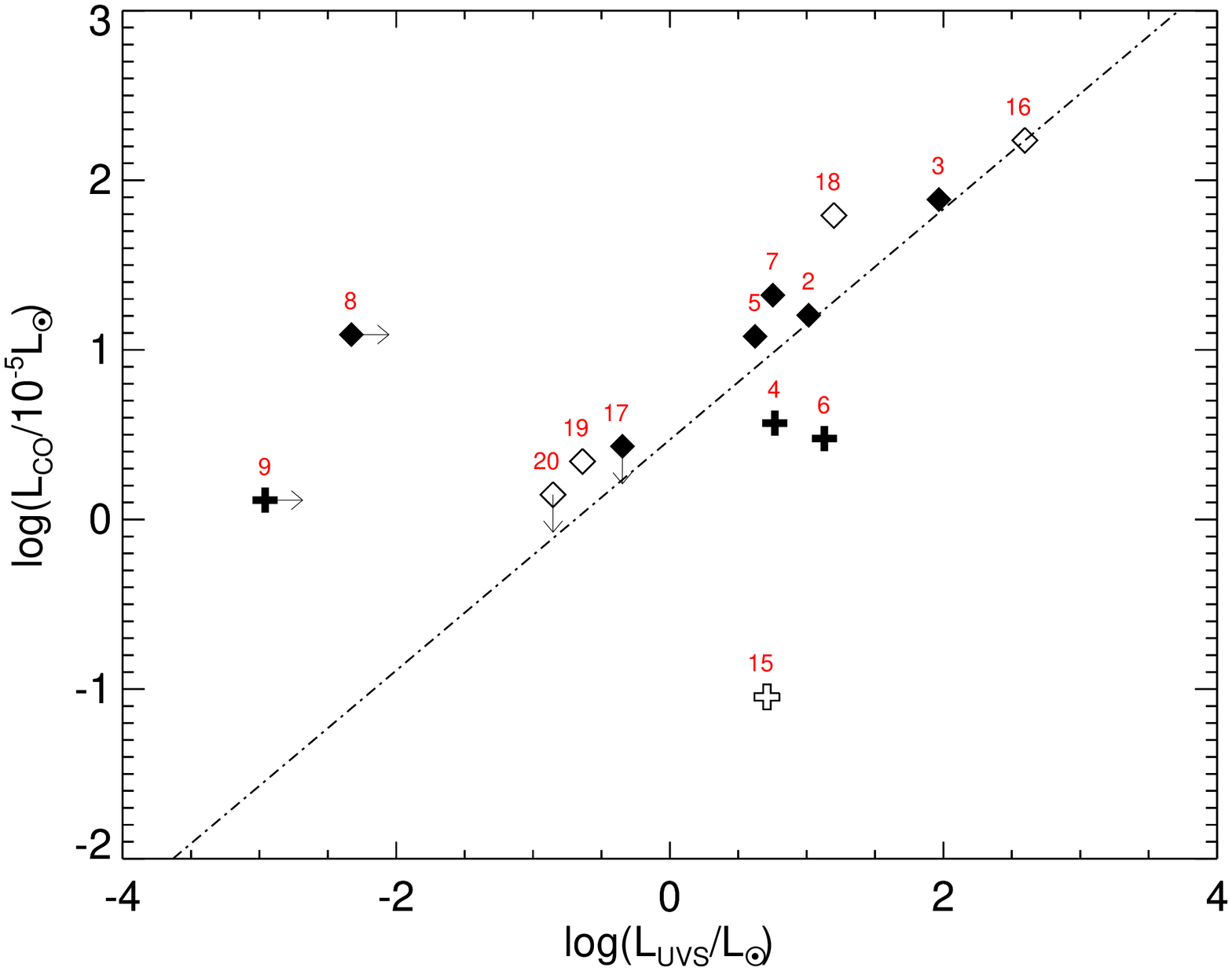} 
\caption{The logarithm of the luminosity of the v=1--0 P30 CO line is plotted versus the
 logarithm of the ultraviolet luminosity integrated from 1300\AA\ to 1850\AA\ 
 (L$\rm_{UVS}$).  The same symbols are repeated from figure 3. 
 The transition disks in the sample 
 are systematically offset from the broader sample. 
 The result of the Buckley-James regression is plotted with the
 dot-dashed line $\rm 0.68log(L_{UVS}/L_{\sun}) + 0.47$ }
\label{ }
\end{center}
\end{figure}

\begin{figure}
\begin{center}
\includegraphics[width=4.5in, angle=-90]{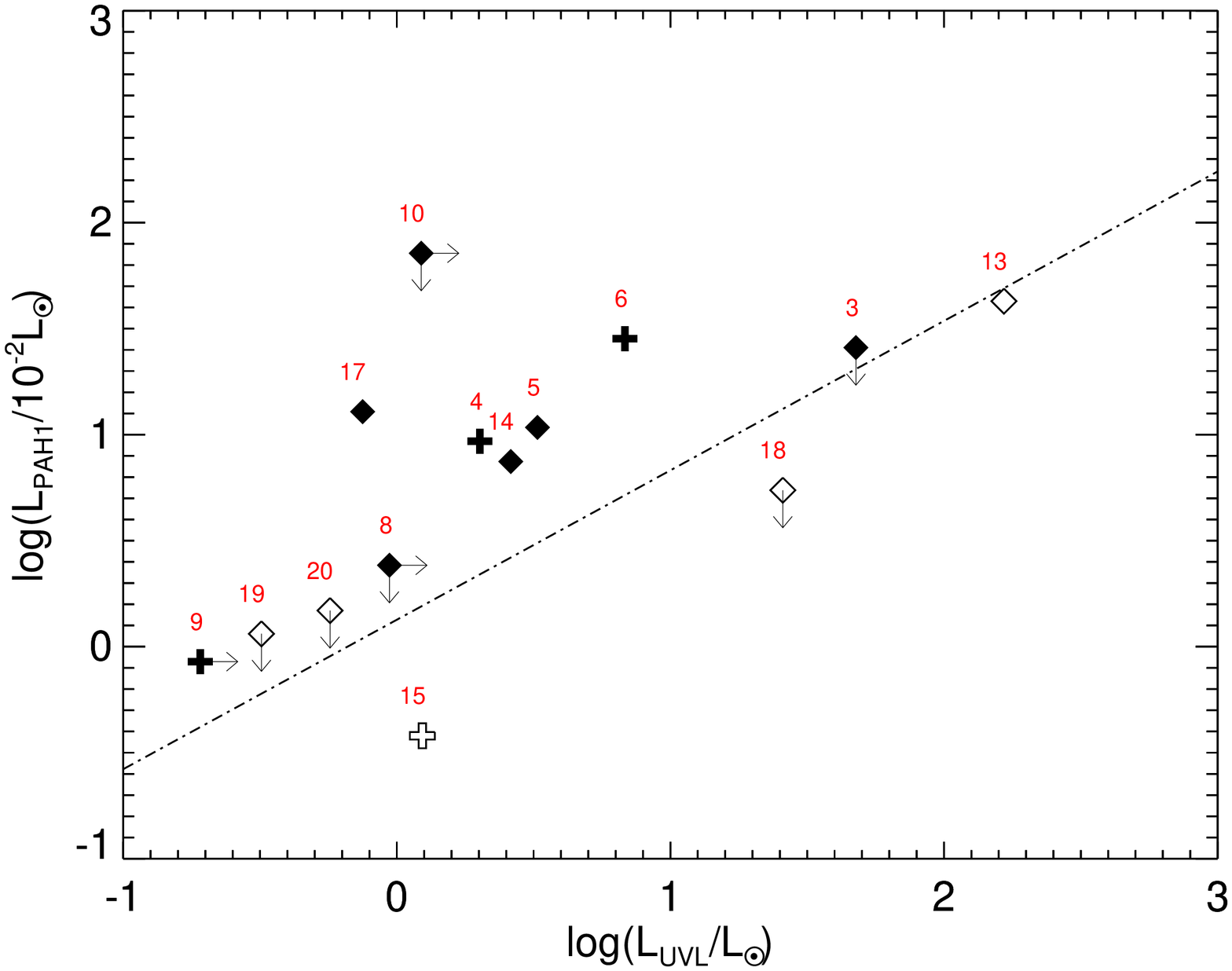} 
\caption{The logarithm of the luminosity of the 7.8$\micron$ PAH line is plotted versus the
 logarithm of the ultraviolet luminosity integrated from 2430\AA\ to 3200\AA\ 
 (L$\rm_{UVL}$). The same symbols are repeated from figure 3. 
 The transition disks in the sample 
 are not systematically offset from the broader sample. 
 The result of the Buckley-James regression is plotted with the
 dot-dashed line $\rm 0.37log(L_{UVL}/L_{\sun}) + 0.61 $ }
\label{ }
\end{center}
\end{figure}

\end{document}